\documentclass{anstrans}

\title{Understanding Chemical Short-Range Order in CoNiV via Mode Analysis}

\author{Jilang Miao$^{*}$ and Miaomiao Jin$^{*}$}

\institute{%
$^{*}$Department of Nuclear Engineering,
Pennsylvania State University, University Park, PA 16802
}

\usepackage{graphicx}
\usepackage{booktabs}
\usepackage{amsmath}
\usepackage{microtype}
\usepackage[section]{placeins}  
\usepackage{stfloats}            

\graphicspath{{fig/}}

\begin{document}
\maketitle
{\let\thefootnote\relax\footnotetext{%
$^\S$Corresponding author: Jilang Miao, \texttt{jlmiao@psu.edu}}}

\section{Introduction}
Multi-principal-element alloys, including CrCoNi and CoNiV, have attracted broad interest as candidate structural materials for nuclear service environments because of their high strength, complex deformation behavior, and potential radiation tolerance. In these chemically concentrated alloys, the local arrangement of elements is not necessarily random. Instead, atoms can exhibit chemical short-range order (SRO), defined by the preference or avoidance of specific element pairs over one or more coordination shells. Such local chemical order can influence defect energetics, dislocation motion, phase stability, and irradiation response~\cite{zhang2020crconi,chen2021direct,lu2025dualscale}.

A commonly used measure of SRO is the Warren--Cowley (WC) parameter~\cite{cowley1950},
\begin{equation}\label{eq}
\alpha_{AB}^{(s)} = 1 - \frac{P^{(s)}(B \mid A)}{c_B},
\end{equation}
where $P^{(s)}(B \mid A)$ is the conditional probability that a neighbor of an $A$ atom in coordination shell $s$ is of type $B$, and $c_B$ is the bulk concentration of element $B$. A negative value of $\alpha_{AB}^{(s)}$ indicates a preference for unlike neighbors, corresponding to chemical ordering, whereas a positive value indicates avoidance of unlike neighbors or a clustering tendency. The random alloy limit is given by $\alpha_{AB}^{(s)} = 0$.

Although the WC parameter provides a clear pairwise definition of SRO, its interpretation becomes difficult in multi-component alloys. For an alloy with $K$ elements in $S$ coordination shells, the SRO state is described by $K(K+1)/2 \times S$ closely related quantities. Reading these values pair by pair can obscure the dominant collective ordering pattern, especially when multiple pair correlations evolve together. A further challenge is convergence. Simulating SRO by atom-swap Monte Carlo (MC) is slow because ordering involves correlated changes in many local environments. Conventional indicators, such as the total energy or radial distribution function (RDF), can appear saturated even when the chemical SRO continues to evolve. A convergence metric based directly on SRO is therefore needed to determine whether the sampled chemical state is stable.

In this work, we use equiatomic CoNiV as a model system to examine how shell-resolved SRO reveals collective ordering modes during atom-swap MC sampling. Starting from the underlying pair statistics used to compute the WC parameters, we apply principal component analysis (PCA) to identify coupled variations across element pairs and coordination shells. We then use a Jensen--Shannon divergence (JSD) lag test to evaluate whether these chemical ordering features remain statistically stable over MC time. This combined analysis helps interpret SRO beyond individual pairwise WC parameters.

\section{Methods}
\subsection{Hybrid MD/MC Simulations}

Hybrid molecular dynamics/Monte Carlo (MD/MC) simulations were performed in LAMMPS \cite{LAMMPS} for an equiatomic CoNiV face-centered cubic (FCC) solid solution containing 2048 atoms ($8\times8\times8$ conventional cells; $L = 28.72$~\AA; $a_0 \approx 3.59$~\AA). Interatomic interactions were described using a level-20 moment tensor potential (MTP) through the Machine Learning Interatomic Potentials (MLIP) package~\cite{shapeev2016mtp,novikov2021mlip,podryabinkin2023mlip3,chen2026chemical}. After box relaxation and a 50~ps isothermal-isobaric (NPT) anneal at 300~K, chemical sampling was performed in the canonical (NVT) ensemble using atom-swap MC moves~\cite{sadigh2012mc}. Swaps among Ni--Co, Co--V, and Ni--V pairs were attempted every 100 MD steps, with 100 attempts per pair type. The production run was continued for 170{,}000 steps, and snapshots were saved regularly, giving 252 snapshots per simulation. Six independent simulations (replicas), differing only in the initial atom random distribution, were used for statistical analysis.

\subsection{Shell-Resolved Pair Statistics}

\begin{figure}[!tb]
\centering
\includegraphics[width=\columnwidth]{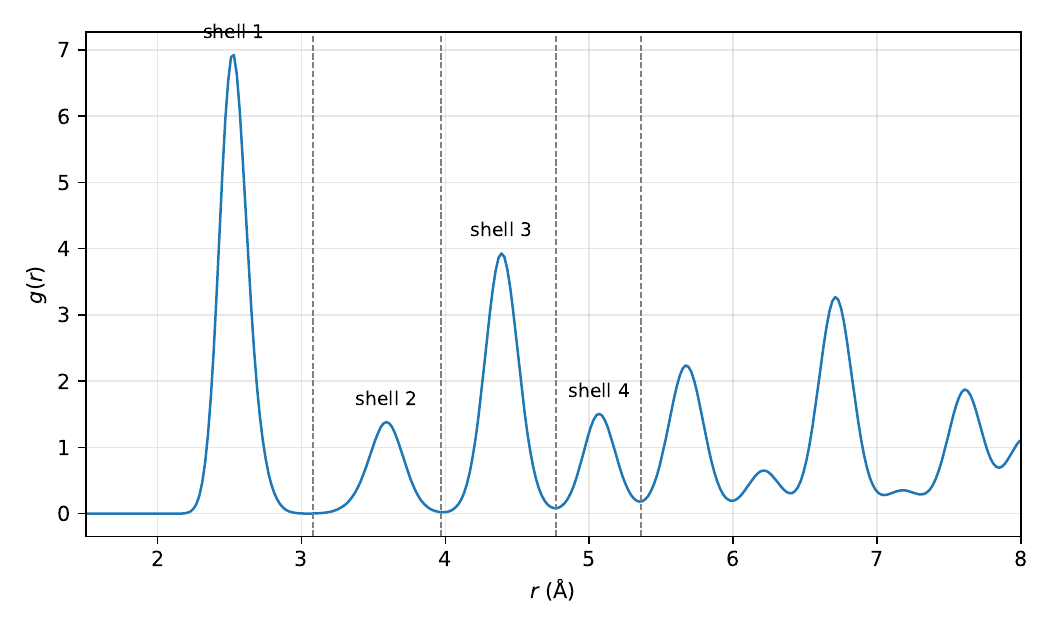}
\caption{Replica- and time-averaged total $g(r)$ of equiatomic CoNiV at
300~K. Dashed lines mark the fixed shell boundaries
$r_{s,\min}$ used for all pair statistics.}
\label{fig:rdf}
\end{figure}

The total radial distribution function $g(r)$ was computed under periodic boundary conditions, as shown in Fig. \ref{fig:rdf}. The resulting minima were used to define the shells. and the corresponding coordination numbers,
$z = [12, 6, 24, 12]$, as expected from the FCC lattice.

Within each shell $s$, we computed two related quantities: the pair (A-B) fractions
$P_{\mathrm{pair}}^{(s)}(A,B)$ and the conditional probabilities
$P^{(s)}(B \mid A)$. For $K=3$ elements and $S=4$ shells, the pair fractions form a 24-dimensional feature vector. Warren--Cowley parameters can be computed using Eq.~\eqref{eq}. The pair fraction vectors were used for PCA and convergence analysis.

\subsection{Stationarity Monitoring by Jensen--Shannon Distance}

The Jensen--Shannon divergence (JSD) between two discrete probability distributions $p$ and $q$ is defined as~\cite{lin1991}
\begin{equation}\label{eq}
\mathrm{JSD}(p,q) = \frac{1}{2}\sum_i p_i \ln\frac{p_i}{m_i}
+ \frac{1}{2}\sum_i q_i \ln\frac{q_i}{m_i},
\end{equation}
where $m = \tfrac{1}{2}(p+q)$. We report the corresponding Jensen--Shannon distance,
$d_{\mathrm{JS}} = \sqrt{\mathrm{JSD}}$~\cite{endres2003}.

To improve statistics, snapshots were grouped into non-overlapping blocks, and block-averaged distributions were compared using $d_{\mathrm{JS}}$. The two distributions tested were the RDF distribution and the shell-resolved pair-fraction vectors, $P_{\mathrm{pair}}$. Two stationarity tests were used. First, neighboring blocks were compared to identify the initial relaxation regime. Second, blocks separated by increasing lag $\ell$ were compared to determine whether the chemical state continued to evolve over longer MC sampling times. If the sampled chemical state is stable, the mean JS distance should increase from a correlated short-lag value and then reach a plateau. Continued growth with lag indicates further evolution of the chemical state.

\subsection{Collective Modes by PCA}
Although the WC parameters provide a direct pairwise measure of SRO, they do not by themselves identify which shell--pair features vary together. PCA was therefore used as a complementary analysis to extract collective modes of variation in the shell-resolved pair statistics. Each PC represents a correlated change across multiple element pairs and coordination shells, allowing the ordering process to be interpreted as a collective mode rather than as isolated changes in individual WC parameters.

Principal component analysis (PCA) was applied to the 24-dimensional $P_{\mathrm{pair}}$ vectors within the post-saturation window. This window was defined as steps $\geq 21{,}000$, chosen after the first-shell $\alpha$ traces had flattened, as discussed later in Sec.~\ref{sec:slow}. For each simulation, the mean $P_{\mathrm{pair}}$ vector was subtracted to remove replica-to-replica offsets. The centered data from all six simulations were then pooled for PCA. The resulting principal components describe correlated fluctuations in shell-resolved pair fractions.

\section{Results}

\subsection{Warren-Cowley Parameters}

\begin{figure}[!tb]
  \centering
  \includegraphics[width=\columnwidth]{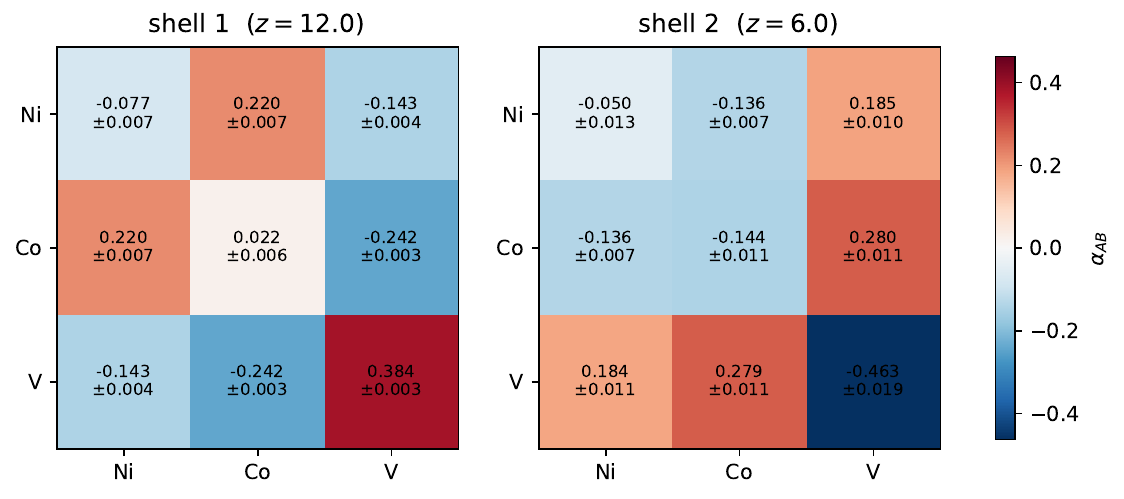}
  \caption{Warren-Cowley parameters $\alpha_{AB}^{(s)}$ of equiatomic
           NiCoV at 300~K for shells 1 and 2 (mean $\pm$ between-replica
           std over six replicas). }
  \label{fig:alpha}
\end{figure}

Figure~\ref{fig:alpha} reports the WC parameters of the first two shells,
averaged over the post-saturation window of each replica and over the six
replicas; the uncertainty is the between-replica standard deviation. The first shell shows strong V--V avoidance, together with preferential Ni--V and Co--V
nearest-neighbor pairing. In the second shell, the V-containing entries
change sign, indicate the opposite preference. Thus, V atoms
avoid one another as nearest neighbors but are enriched at second-neighbor
positions. This shell-dependent sign change is consistent with partial
L1$_2$-like ordering, in which V preferentially occupies one sublattice
relative to Ni and Co. Experimentally, an L1$_1$-like CSRO motif has been reported for
CoNiV~\cite{chen2022motif}, and the first-shell pair preferences
are consistent with prior work~\cite{chen2021direct,kostiuchenko2020sro}.

The between-replica scatter is small, with a maximum standard deviation of
0.019. However, as shown below, the first-shell values are already stable
within the post-saturation window, whereas the second-shell WC entries
continue to evolve slowly. The reported second-shell magnitudes should
therefore be interpreted as lower bounds on the fully equilibrated
ordering.

\subsection{Slow Chemical Ordering}
\label{sec:slow}

\begin{figure}[!tb]
  \centering
  \includegraphics[width=\columnwidth]{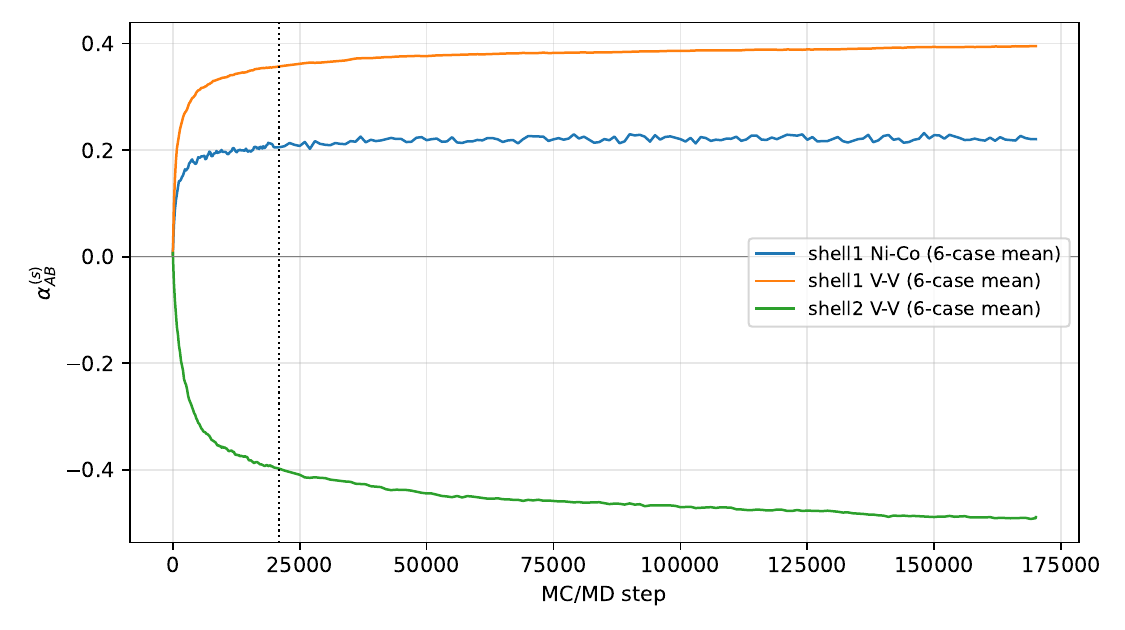}
  \caption{Six-replica mean of selected WC parameters over the full
           trajectory.  }
  \label{fig:alphaevol}
\end{figure}

The lag-JSD test was applied to the post-saturation window, defined as steps $\geq 21{,}000$. As shown in Fig.~\ref{fig:lag}, the RDF lag-JSD remains nearly flat, indicating that the pair-distance distribution is stable. In contrast, the $P_{\mathrm{pair}}$ lag-JSD increases with lag in all six replicas, with a last-to-first-lag ratio of 2.6--3.2. This shows that the chemical pair distribution continues to evolve even after the RDF has stabilized. The slow chemical evolution is also visible in the WC curves
(Fig.~\ref{fig:alphaevol}). The first-shell entries flatten within
$\sim$20{,}000 steps, while $\alpha_{VV}^{(2)}$ continues to decrease
through the full trajectory, without saturation at the end of simulation.

\begin{figure}[!tb]
\centering
\includegraphics[width=\columnwidth]{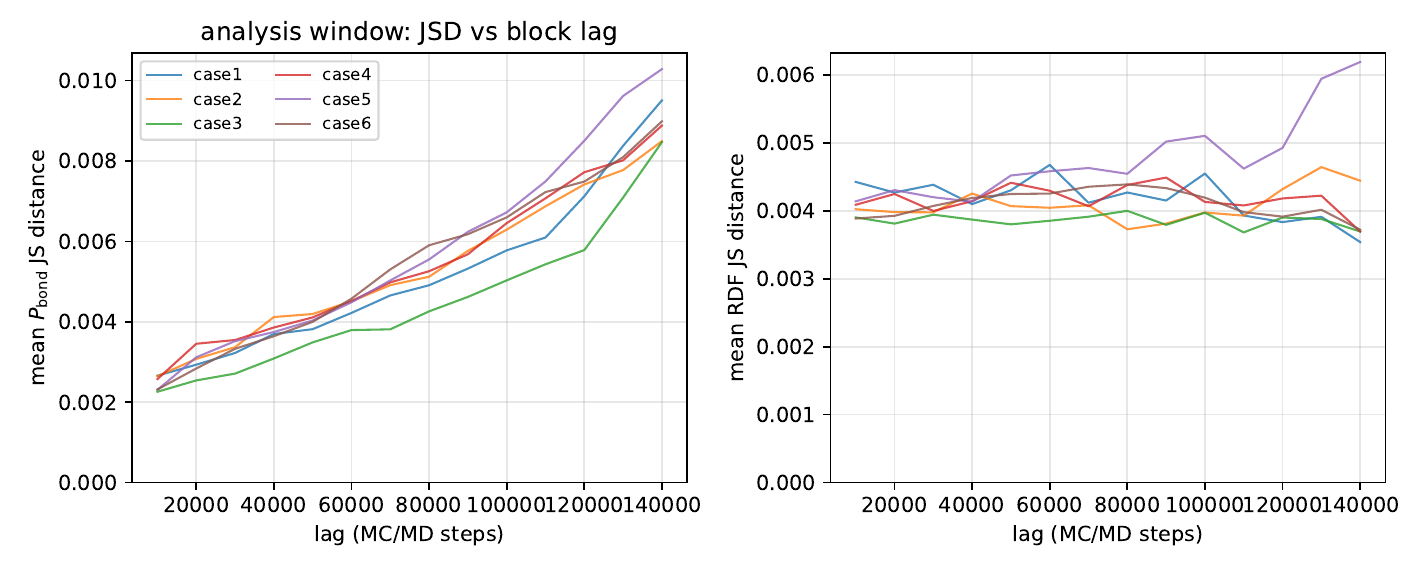}
\caption{Mean JS distance between blocks separated by increasing lag
in the post-saturation window.
The $P_{\mathrm{pair}}$ distance increases with lag in all six
replicas, indicating continued chemical ordering, whereas the
RDF distance remains nearly flat.}
\label{fig}
\end{figure}

\subsection{Collective Modes}

PCA was applied to the per-replica-centered $P_{\mathrm{pair}}$ vectors in the post-saturation window to identify the dominant correlated variations among shell-resolved pair fractions. The first four explained variance ratios are $[0.289, 0.218, 0.154, 0.103]$, and the first three modes together account for 66\% of the variance and are therefore used to describe the main collective ordering modes.

\begin{figure}[!tb]
  \centering
  \includegraphics[width=\columnwidth]{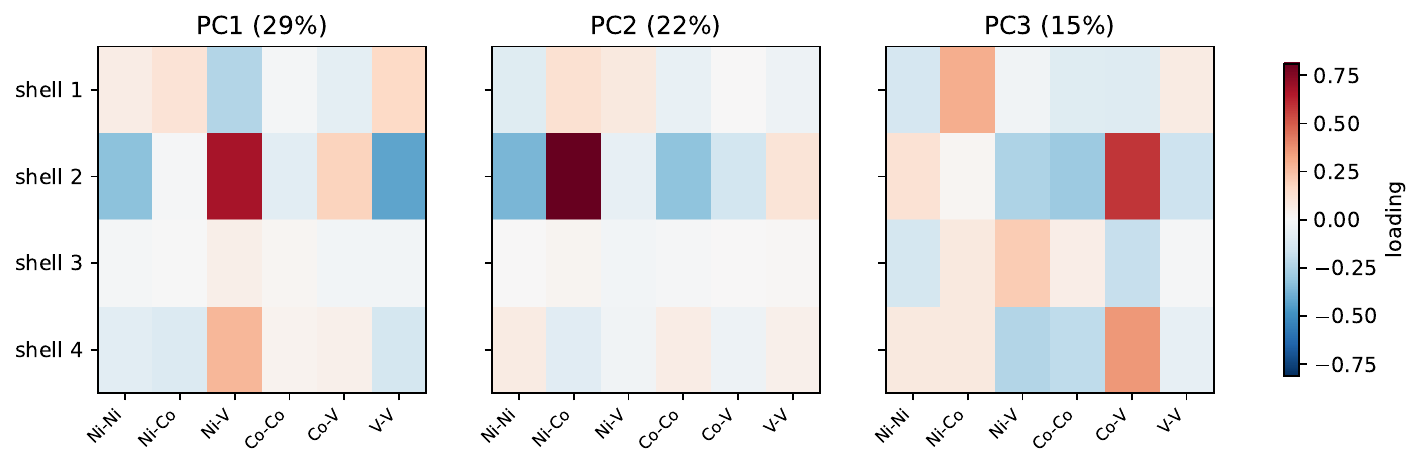}
  \caption{Entries of the three leading PCs, reshaped so that rows correspond to shells and columns correspond to element pairs.}
  \label{fig:loadings}
\end{figure}

To quantify where each mode is localized, the contribution of shell $s$ was computed as the sum of the squared PC-vector entries over the six pair features in that shell. They measure the variance carried by each shell within the sampled window. PC1 accounts for 29\% of the fluctuation variance and is dominated by shell 2, with a shell weight of 0.78, together with a weaker shell-4 contribution of 0.12. In shell 2, the largest entries highlight V-containing pairs, mainly Ni--V $(+0.67)$ and Co--V $(+0.18)$, against V--V $(-0.43)$ and Ni--Ni $(-0.33)$ pairs. The same sign pattern appears more weakly in shell 4. With the sign convention of Fig.~\ref{fig:loadings}, motion along $-$PC1 corresponds to increasing second-shell V--V order.

This mode manifests the nature of the L1$_2$-like SRO. Shells 2 $(a_0,\langle100\rangle)$ and 4 $(a_0\sqrt{2},\langle110\rangle)$ connect sites on the simple-cubic V sublattice of the L1$2$ reference structure, for which the ideal reference gives WC $\alpha{VV}=-3$ in both shells. Their coupled contribution to PC1 indicates that the dominant fluctuation mode is a coherent V-sublattice ordering mode. At the same time, shell 1 contributes weakly to PC1 because the first-shell SRO has already stabilized by the post-relaxation window. By comparison, shell 3 remains weak in both the WC parameters and the PC coefficients. The observed order is therefore not a  developed L1$_2$ pattern across all coordination shells. It is better described as a robust, short-range L1$_2$-like motif dominated by second-shell V ordering. PC1 is therefore interpreted as a coherent V-sublattice ordering mode, measuring the instantaneous degree of L1$_2$-like short-range order across shells 2 and 4.

\begin{figure}[!tb]
\centering
\includegraphics[width=\columnwidth]{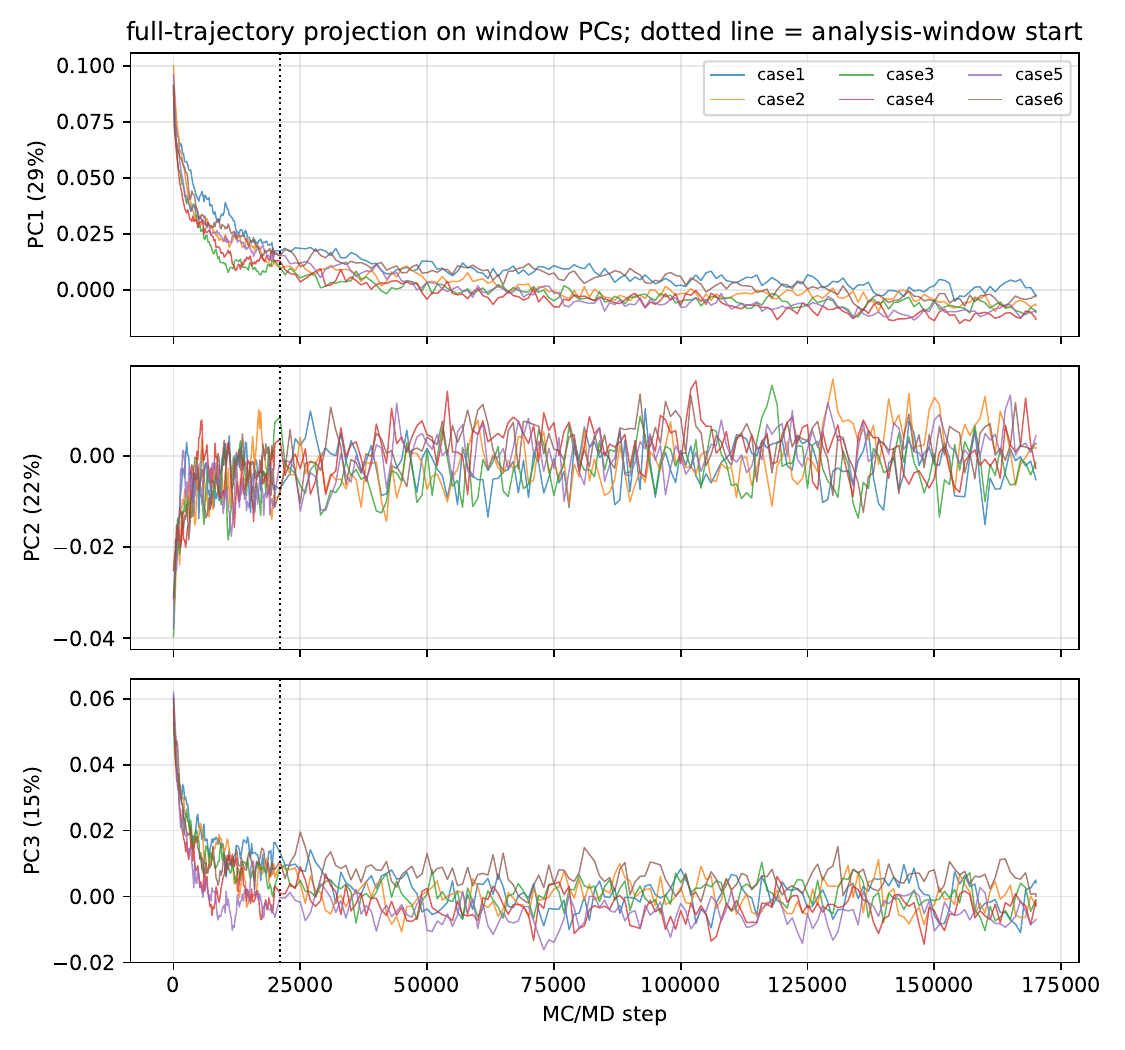}
\caption{PC scores obtained by projecting the full trajectories onto the
three leading PCs. The dotted line marks the start of the
post-relaxation window at step 21{,}000. PC1 continues to evolve
throughout the run, whereas PC2 and PC3 fluctuate around zero.}
\label{fig:evol}
\end{figure}

PC2 accounts for 22\% of the fluctuation variance and is almost entirely localized in shell 2, with a weight of 0.94. Its largest entries highlight Ni--Co $(+0.81)$ against Ni--Ni $(-0.37)$ and Co--Co $(-0.33)$, while all V-containing entries are below 0.16 in magnitude. PC2 therefore represents a Ni--Co redistribution mode at nearly fixed V order. PC3, which accounts for 15\% of the fluctuation variance, contains mixed Co--V contributions across shells 2 and 4. These modes show that Ni/Co redistribution and V-sublattice ordering are not simply one combined process, but appear as distinguishable collective variations in the pair-statistics space.
The dynamical relevance of these modes was examined by projecting the full trajectories onto the PCs obtained from the post-saturation window. As shown in Fig.~\ref{fig:evol}, all replicas first move rapidly away from the random initial state and then continue to evolve slowly along $-$PC1 throughout the simulation. In contrast, PC2 and PC3 fluctuate around zero without a sustained trend. PC1 therefore captures both the leading fluctuation mode in the post-saturation window and the direction of the continued second-shell V ordering identified in Sec.~\ref{sec:slow}.

Overall, the mode analysis adds three insights beyond the WC parameters alone. First, the dominant SRO variation is shell-selective and controlled mainly by second-shell V ordering. Second, shell 3 does not participate strongly in the L1$_2$-like motif, supporting a short-range rather than fully developed L1$_2$ interpretation. Third, Ni--Co redistribution forms a separate collective mode from V-sublattice ordering. Together, these results support a robust short-range L1$_2$-like ordering tendency in CoNiV.

\section{Conclusions}

We developed a shell-resolved pair-statistics analysis to quantify chemical short-range order in equiatomic CoNiV and assess its convergence during hybrid MD/MC sampling. At 300~K, the WC parameters show first-shell V--V avoidance and second-shell V--V enhancement, consistent with an L1$_2$-like ordering tendency. Notably, the ordering is shell-selective, where PCA shows that shells 2 and 4 contribute coherently to a V-sublattice-like ordering mode, while shell 3 remains weak. The observed SRO is therefore better described as a robust short-range L1$_2$-like motif. The convergence and mode analyses further show that first- and second-shell SRO evolve on different timescales. First-shell WC parameters flatten much faster than the higher shells. RDF-based diagnostics indicate a stable pair-distance structure, but the lag-JSD test on $P_{\mathrm{pair}}$ detects continued chemical evolution, showing that structural convergence does not guarantee SRO convergence. PCA identifies this slow evolution as a V-sublattice ordering and Ni--Co redistribution. The mode analysis identifies which shell--pair features evolve together and provides a practical diagnostic for chemical equilibration in multi-principal-element alloys.

\section*{Acknowledgments}

The authors thank the Penn State Institute for Computational and Data
Sciences (ICDS) for the high-performance computing resources used in this
work, and gratefully acknowledge support from the ICDS Rising Researchers
Award.

\bibliographystyle{ans}
{\footnotesize
\bibliography{refs}

@article{chen2026chemical,
  title={Chemical Short-Range Order Regulates Hydrogen Energetics and Hydrogen-Dislocation Interactions in CoNiV},
  author={Chen, Beihan and Ahmed, Dalia Sayed and Yang, Yang and Jin, Miaomiao},
  journal={arXiv preprint arXiv:2604.05352},
  year={2026}
}

@Article{LAMMPS,
  author = "A. P. Thompson and H. M. Aktulga and R. Berger and 
     D. S. Bolintineanu and W. M. Brown and P. S. Crozier and
     P. J. in 't Veld and A. Kohlmeyer and S. G. Moore and T. D. Nguyen and
     R. Shan and M. J. Stevens and J. Tranchida and C. Trott and S. J. Plimpton",
  title = "{LAMMPS} - a flexible simulation tool for
     particle-based materials modeling at the 
     atomic, meso, and continuum scales",
  journal = "Comp. Phys. Comm.",
  volume =  "271",
  pages =   "108171",
  year =    "2022",
  doi = "10.1016/j.cpc.2021.108171"
}

@article{cowley1950,
  author  = {Cowley, J. M.},
  title   = {An Approximate Theory of Order in Alloys},
  journal = {Physical Review},
  volume  = {77},
  pages   = {669--675},
  year    = {1950},
  doi     = {10.1103/PhysRev.77.669}
}

@article{chen2021direct,
  author  = {Chen, Xuefei and Wang, Qi and Cheng, Zhiying and Zhu, Mingliu and
             Zhou, Hao and Jiang, Ping and Zhou, Lingling and Xue, Qiqi and
             Yuan, Fuping and Zhu, Jing and Wu, Xiaolei and Ma, En},
  title   = {Direct observation of chemical short-range order in a
             medium-entropy alloy},
  journal = {Nature},
  volume  = {592},
  pages   = {712--716},
  year    = {2021},
  doi     = {10.1038/s41586-021-03428-z}
}

@article{chen2022motif,
  author  = {Chen, Xuefei and Yuan, Fuping and Zhou, Hao and Wu, Xiaolei},
  title   = {Structure motif of chemical short-range order in a
             medium-entropy alloy},
  journal = {Materials Research Letters},
  volume  = {10},
  pages   = {149--155},
  year    = {2022},
  doi     = {10.1080/21663831.2022.2029607}
}

@article{kostiuchenko2020sro,
  author  = {Kostiuchenko, Tatiana and Ruban, Andrei V. and Neugebauer,
             J{\"o}rg and Shapeev, Alexander and K{\"o}rmann, Fritz},
  title   = {Short-range order in face-centered cubic {VCoNi} alloys},
  journal = {Physical Review Materials},
  volume  = {4},
  pages   = {113802},
  year    = {2020},
  doi     = {10.1103/PhysRevMaterials.4.113802}
}

@article{zhang2020crconi,
  author  = {Zhang, Ruopeng and Zhao, Shiteng and Ding, Jun and Chong, Yan and
             Jia, Tao and Ophus, Colin and Asta, Mark and Ritchie, Robert O.
             and Minor, Andrew M.},
  title   = {Short-range order and its impact on the {CrCoNi} medium-entropy
             alloy},
  journal = {Nature},
  volume  = {581},
  pages   = {283--287},
  year    = {2020},
  doi     = {10.1038/s41586-020-2275-z}
}

@article{lu2025dualscale,
  author  = {Lu, Tiwen and Sun, Binhan and Li, Yue and Dai, Sheng and Yao,
             Ning and Li, Wenbo and Dong, Xizhen and Chen, Xiyu and Niu,
             Jiacheng and Ye, Fan and Kwiatkowski da Silva, Alisson and Zhu,
             Shuya and Xie, Yu and Yang, Xiaofeng and Deng, Sihao and Tan,
             Jianping and Li, Zhiming and Ponge, Dirk and He, Lunhua and
             Zhang, Xian-Cheng and Raabe, Dierk and Tu, Shan-Tung},
  title   = {Dual-scale chemical ordering for cryogenic properties in
             {CoNiV}-based alloys},
  journal = {Nature},
  volume  = {645},
  pages   = {385--391},
  year    = {2025},
  doi     = {10.1038/s41586-025-09458-1}
}

@article{shapeev2016mtp,
  author  = {Shapeev, Alexander V.},
  title   = {Moment Tensor Potentials: A Class of Systematically Improvable
             Interatomic Potentials},
  journal = {Multiscale Modeling \& Simulation},
  volume  = {14},
  pages   = {1153--1173},
  year    = {2016},
  doi     = {10.1137/15M1054183}
}

@article{novikov2021mlip,
  author  = {Novikov, Ivan S. and Gubaev, Konstantin and Podryabinkin,
             Evgeny V. and Shapeev, Alexander V.},
  title   = {The {MLIP} package: moment tensor potentials with {MPI} and
             active learning},
  journal = {Machine Learning: Science and Technology},
  volume  = {2},
  pages   = {025002},
  year    = {2021},
  doi     = {10.1088/2632-2153/abc9fe}
}

@article{podryabinkin2023mlip3,
  author  = {Podryabinkin, Evgeny and Garifullin, Kamil and Shapeev,
             Alexander and Novikov, Ivan},
  title   = {{MLIP-3}: Active learning on atomic environments with moment
             tensor potentials},
  journal = {The Journal of Chemical Physics},
  volume  = {159},
  pages   = {084112},
  year    = {2023},
  doi     = {10.1063/5.0155887}
}

@article{sadigh2012mc,
  author  = {Sadigh, Babak and Erhart, Paul and Stukowski, Alexander and
             Caro, Alfredo and Martinez, Enrique and Zepeda-Ruiz, Luis},
  title   = {Scalable parallel Monte Carlo algorithm for atomistic
             simulations of precipitation in alloys},
  journal = {Physical Review B},
  volume  = {85},
  pages   = {184203},
  year    = {2012},
  doi     = {10.1103/PhysRevB.85.184203}
}

@article{lin1991,
  author  = {Lin, Jianhua},
  title   = {Divergence measures based on the {Shannon} entropy},
  journal = {IEEE Transactions on Information Theory},
  volume  = {37},
  pages   = {145--151},
  year    = {1991},
  doi     = {10.1109/18.61115}
}

@article{endres2003,
  author  = {Endres, Dominik M. and Schindelin, Johannes E.},
  title   = {A new metric for probability distributions},
  journal = {IEEE Transactions on Information Theory},
  volume  = {49},
  pages   = {1858--1860},
  year    = {2003},
  doi     = {10.1109/TIT.2003.813506}
}
}

\end{document}